%% file: Stelter_MacroSlicer_SPIE_arxiv_submission.tex
\documentclass[a4paper]{spie} 

\usepackage{graphicx} 
\usepackage{color} 
\usepackage{xfrac} 
\usepackage{tabularx} 
\usepackage{booktabs} 

\usepackage{caption} 

\usepackage{comment} 

\graphicspath{{./Figures/}
{./Figures/TikZ}
{./TikZ-Figures/}
{./Python-Figures/}
}

\newcommand{\til}{$\sim$}
\newcommand{\farcsec}{\mbox{\ensuremath{\,.\!\!^{\prime\prime}}}}
\newcommand{\arcsec}{\mbox{\ensuremath{\,\,\!\!^{\prime\prime}}}}
\newcommand{\fiftyeed}{50\%EED}

\title{Optical and Opto-Mechanical Design of a Novel ``Macro'' Image Slicer for the MIRADAS Instrument}

\author{R.~Deno Stelter\supit{a} and Stephen S.~Eikenberry\supit{b}
\skiplinehalf
\supit{a}Center for Adaptive Optics, 1156 High Street, UC Santa Cruz, Santa Cruz, CA 95063 USA \\
\supit{b}University of Florida, 211 Bryant Space Center, Gainesville, FL 32611 USA 
}

\authorinfo{Further author information: (Send correspondence to R.~Deno Stelter)\\
R.~Deno Stelter: Email: deno2ucolick.org, Office: +1 831 459-5780 \\
Stephen S.~Eikenberry: Email: eiken@ufl.edu, Office: +1 352 294 1833}

\begin{document}
\maketitle

\begin{abstract} 
We present the innovative macro-slicer optical and opto-mechanical designs for the third-generation Mid-resolution InfraReD Astronomical Spectrograph (MIRADAS) instrument for the 10.4m Gran Telescopio Canarias (GTC) in the 1-2.5 $\mu$m bandpass. 
MIRADAS uses up to 12 cryogenic, fully steerable probes to select simultaneous targets in a 5 arcminute field of view. 
The spectrograph module is a cross-dispersed echelle spectrograph. 
The macro-slicer is effectively a stack of six advanced image slicer Integral Field Units (IFUs) such as FRIDA or FISICA, and like other IFUs designed and built at the University of Florida by our group, uses a `bolt-and-go' approach to minimize the difficulty in alignment and maximize robustness. 
Like other advanced image slicer IFUs, there are three sets of mirrors that work together to geometrically rearrange the loosely packed inputs from the probe arms into a tightly packed pseudo-slit. 
The macro-slicer also passively keeps the spectral resolution of MIRADAS fixed at $R>20,000$ in seeing from 1.2 arcseconds down to 0.4 arcseconds, (typical observing conditions at GTC).  
\end{abstract}


\section{Introduction}
The GTC is currently the world's largest steerable optical/infrared telescope and is situated on the island of La Palma in the Canary Islands.
MIRADAS\cite{miradas10,miradas12,miradas16} is designed to occupy one of the Folded Cassegrain focal positions.

This paper describes the optical and mechanical design of the MIRADAS macro-slicer.
The macro-slicer is made of 6 co-mounted modules that are effectively 6 Content-style advanced image slicers\cite{ais98} working in parallel to accept 12 disparate inputs and create a staggered pseudo-slit suited to match the MIRADAS spectrograph module.
By themselves, the macro-slicer modules are advanced image slicers with three powered slices apiece; each slice is geometrically rearranged end-to-end by the three pupil mirrors, (each at or near the pupil plane created by its slice) and re-imaged onto the field mirrors (which are co-located at the image plane generated by the slicer and pupil mirrors).
The field mirrors are un-powered fold mirrors and redirect the light into the spectrograph module.

\section{MIRADAS Overview}
\label{sec:miOverview}
The MIRADAS instrument for the GTC is the third-generation multi-object near-infrared spectrograph with R$>$20 000 over the 1--2.5 $\mu$m bandpass.
MIRADAS can deploy up to 12 cryogenic probe arms\cite{probearms14} simultaneously with pick-off mirrors, each feeding a 3.7 arcsec x 1.2 arcsec field of view to the spectrograph.
The spectrograph input optics also include a “slit slicer” (henceforth referred to as the macro-slicer) which reformats each probe field into 3 end-to-end slices of a fixed 3.7 arcsec x 0.4 arcsec format -- combining the advantages of minimal slit losses in any seeing conditions better than 1.2 arcsec, while at the same time providing some, albeit limited, two-dimensional spatial resolution.
Figure~\ref{fig:MI_blockDiagram} shows a block schematic of the instrument, highlighting the three optical relays or blocks: 1) the probe arms; 2) the macro-slicer; and 3) the spectrograph module.
Figure~\ref{fig:MI_focalPlaneEvolution} shows the corresponding focal planes for each of the blocks.
Mechanically, the probe arms sit on a circular bench which is separated from a circular bulkhead by a cylindrical ring \til175 mm in height.
The bulkhead provides a mounting surface for several of the spectrograph optics.
The volume between the bulkhead and the probe arm bench (called the `fore-volume') is illuminated only by the probe arms; the probe arm bench and bulkhead provide excellent baffling as well as mechanical mounting surfaces for the probe arms, and macro-slicer \& spectrograph module, respectively.

\begin{figure}[htp]
\centering
  \includegraphics[width=0.6\textwidth]{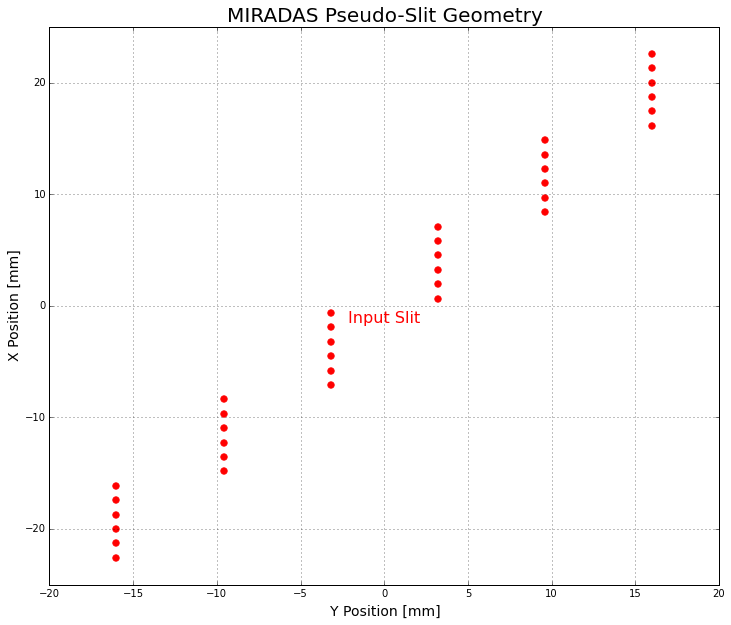}
  \caption{Schematic view of a section of the ``sliced'' probe outputs forming the pseudo-longslit at the third focal plane (also the spectrograph input slit).
    Of note is the offset columns of six rows each -- this is the `stagger' mentioned in the text.
    Each of the dots represents the center of one field mirror.}
\label{fig:MI-SlitGeometry}
\end{figure}

The 12 probe arms each patrol a $30^{\circ}$ `slice of sky pie' and relay light from the first image plane to the entrance image plane of the macro-slicer; the probe arms are functionally positionable periscopes with a fixed optical path length.
The macro-slicer relays the light from its disparate inputs (eg, the probe arms) and geometrically rearranges it from a sparsely-filled, \til200 mm long, 1.2 arcsec wide pseudo-slit into a tightly packed, staggered, \til100 mm long, 0.4 arcsec wide pseudo-slit (see Figure~\ref{fig:MI-SlitGeometry}).
This rearranged pseudo-slit is then relayed into the spectrograph module, which is a high-resolution cross-dispersed echelle spectrograph.

The staggered pseudo-slit is a design choice that allows the spectral orders to fill the detector -- if the pseudo-slit were not staggered, the spectral orders would lie at a $\sim 30^{\circ}$ angle and thus fill only \sfrac{1}{2} of the detector.
Figure~\ref{fig:MI_focalPlaneEvolution} shows that the individual spectral lines are tilted by $\sim 30^{\circ}$ but the spectral orders are close to horizontal.
In order to generate the stagger (the so-called `canonical stagger'), the macro-slicer modules are independently positioned, although they remain co-planar with each other.
The pupil mirrors are also staggered so that the pupil plane is coincident with the pupil mirror.
This allows for MIRADAS to use AB on-source nodding in order to remove telluric lines in the spectra.

\begin{minipage}[b]{0.48\textwidth}
	\centering
	  \includegraphics[width=\textwidth]{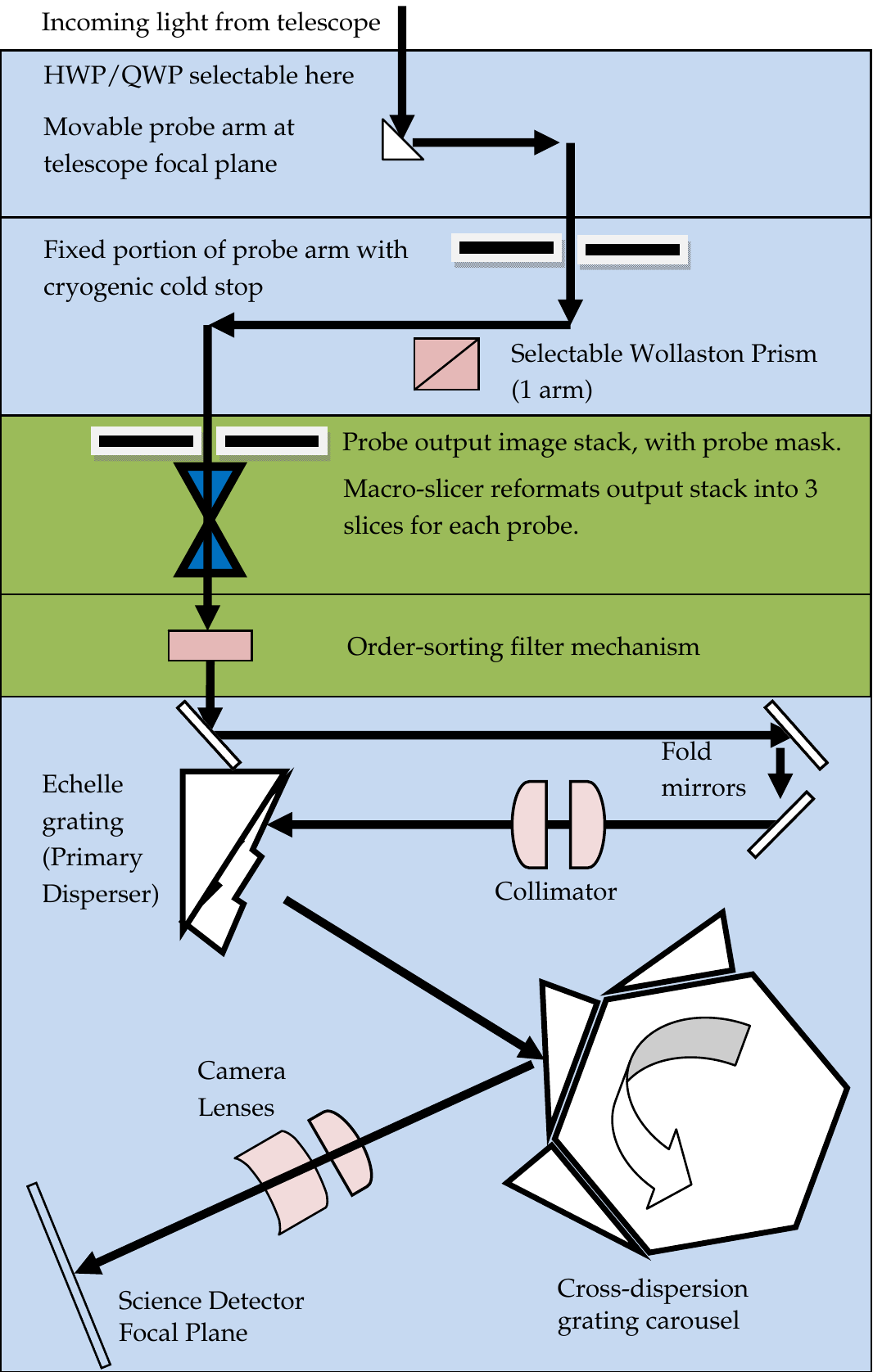}
	  \captionof{figure}{\small{MIRADAS Optical Concept Block Diagram. 
       Incoming light from the telescope enters the instrument at the top of the figure and is selected and folded by the probe arms in the top two blue-shaded regions. 
	    The probe outputs are stacked, sliced, and filtered in the two middle green-shaded regions, after which it enters the cross-dispersed echelle spectrograph module, as represented in the lower blue-shaded region.}}
	\label{fig:MI_blockDiagram}
\end{minipage}
~~~%
\begin{minipage}[b]{0.48\textwidth}
	\centering
	  \includegraphics[width=\textwidth]{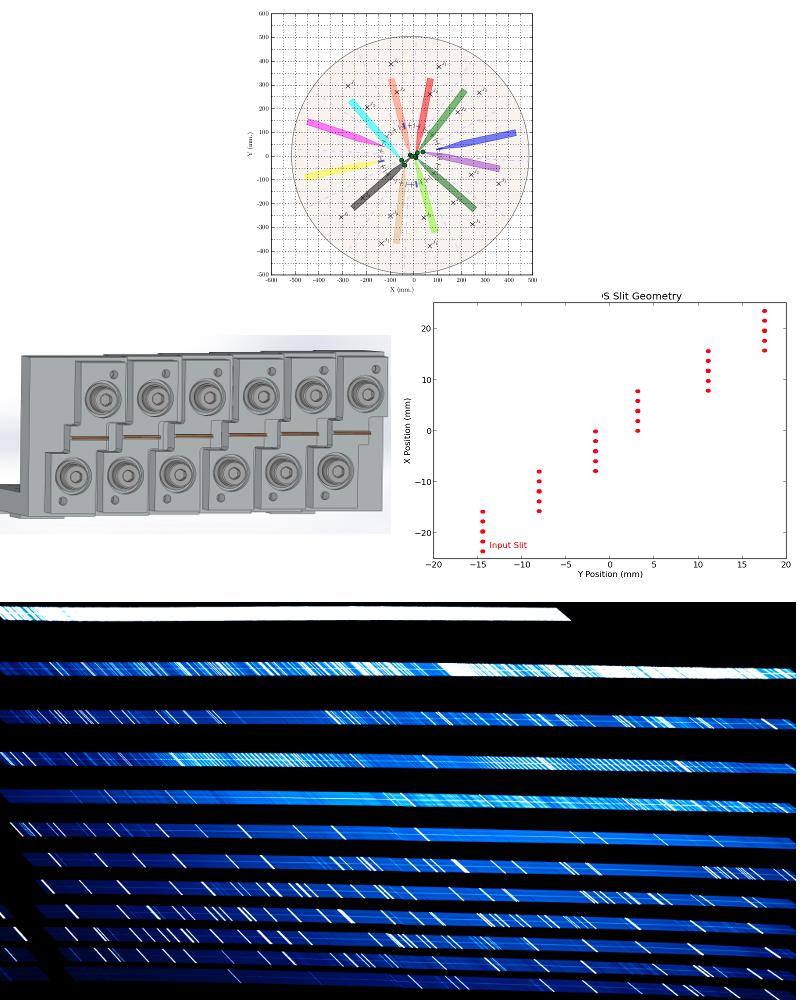}
	  \captionof{figure}{\small{
	  This figure shows the `evolution' of the focal planes of MIRADAS.
	  \textbf{Top:} The first focal plane is just above the probe arm pick-off mirrors, shown here top-down.
	  The inner hashed circle is the 5 arcmin field of regard for MIRADAS.
	  \textbf{Middle left:} The second focal plane is at the macro-slicer input (where the slicer mirrors live).
	  \textbf{Middle right:} The third focal plane is at the macro-slicer output (where the field mirrors live).
	  This focal plane also acts as the input for the spectrograph camera.
	  \textbf{Bottom:} The final focal plane is at the detector.
	  The cross-dispersed spectrum of a single object from 1.25 -- 2.5 $\mu$m, made with our data simulator.
	  }}
	\label{fig:MI_focalPlaneEvolution}
\end{minipage}

\section{Macro-Slicer Overview \& Configuration}
\label{sec:macroSlicerOverview}
Optically, the task of the macro-slicer is to accept the light beams from the probe arms, rearrange them into a tightly-packed pseudo-slit, and relay it into the spectrograph module while minimizing the introduction of additional optical aberrations as much as possible, all the while keeping the throughput high by minimizing the number of mirrors the light from any one probe arm `sees.'
Table~\ref{tab:opticalPrecription} lists the mirrors that make up the macro-slicer and their properties.
Figure~\ref{fig:macroSlicerPics} shows the mechanical design of the macro-slicer assembly on its IFU bench and a prototype of the pupil mirrors.

\begin{minipage}[c]{0.48\textwidth}
	\centering
	  \includegraphics[width=\textwidth]{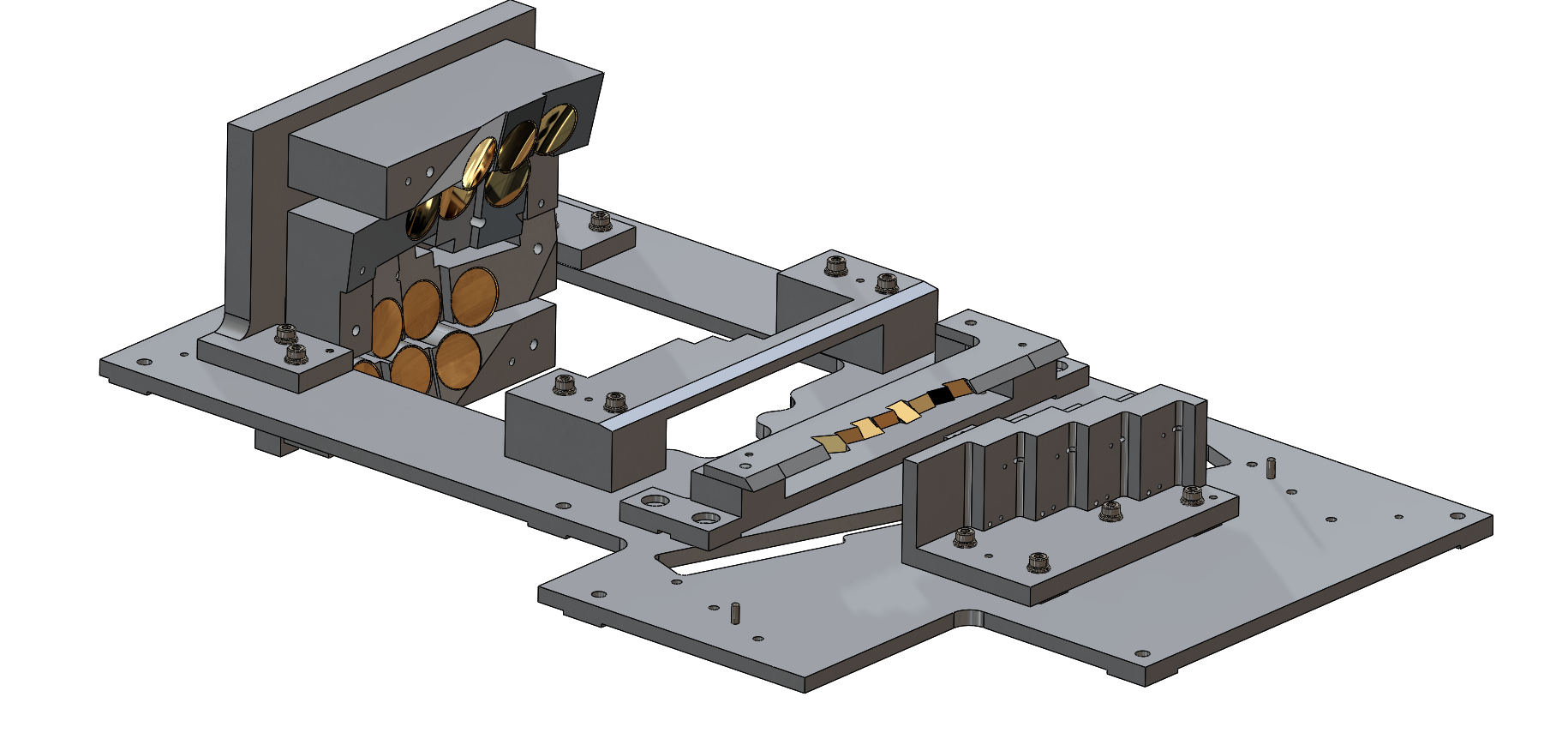}
\end{minipage}
~~
\begin{minipage}[c]{0.48\textwidth}
	\centering
	  \includegraphics[width=\textwidth]{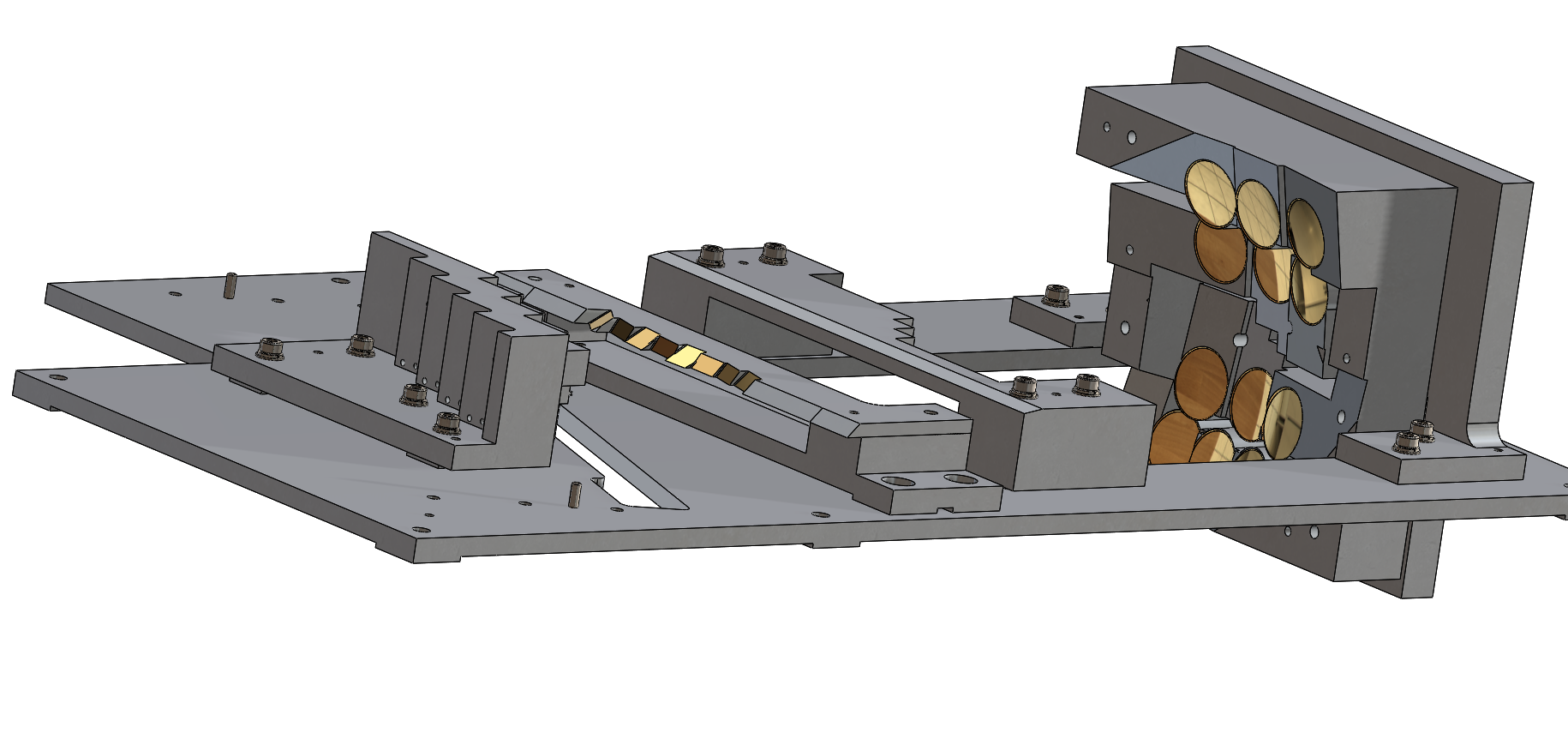}
\end{minipage}
\\
\begin{center}
\begin{minipage}[c]{0.98\textwidth}
	\centering
	  \includegraphics[width=0.89\textwidth]{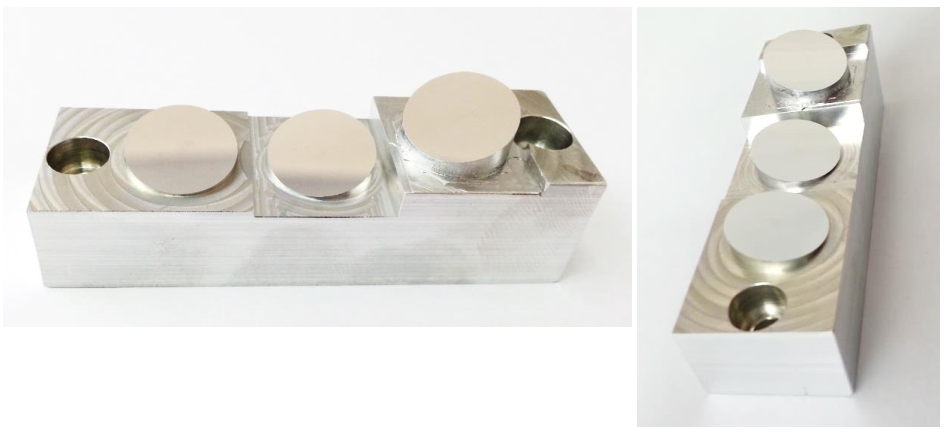}

	  \captionof{figure}{\small{Several renderings of the MIRADAS macro-slicer assembly.
	  \textbf{Top:} The MIRADAS macro-slicer assembly from two isometric views; this shows the macro-slicer configuration for 8 probe arms.
	  The mechanical design of the pupil and slicer mirrors has evolved during consultation with the manufacturer, Durham Precision Optics.
	  The decker mechanism is not shown, but the smaller, rectangular cut-out marks where the decker sits. 
	  \textbf{Bottom:} A prototype of the MIRADAS pupil mirrors (figure courtesy C. Bourgenot, Durham Precision Optics).}} 
\label{fig:macroSlicerPics}
\end{minipage}
\end{center}

\begin{table}
	\centering
	\caption{\small{Overview for macro-slicer components.
		Here we describe the mirrors for each component of the macro-slicer module, beginning with the beam combiner mirrors.
		In total there are 84 mirrors, 36 of them powered, that make up the macro-slicer module; this does not include the aft sytem, or fixed portion of the probe arms.
		We show the nominal values of the `radii of curvature' and `thickness' for the slicer and pupil mirrors; the optical prescription has been optimized on a slitlet-by-slitlet basis as discussed in \S\ref{sec:optimize}.
		The 'f-speed' of each component is the beam speed at the listed component.
		`Plate Scale' is given for those components at an image plane.
		`Diameter/Length' is given for each component; those which are circular do not have any values in the `Width' column. 
		`Thickness' here refers to the track length from the center of the component to the center of the next.
		The beam combiner mirrors are not staggered and are on average 50 mm from the slicer modules, which are staggered in the canonical fashion.
}}%
\label{tab:opticalPrecription}
	\input{opticalPrescription}

\end{table}

Mechanically, the macro-slicer must fit in a tightly-packed volume without interfering with other components.
As demonstrated in Figure~\ref{fig:MI_blockDiagram}, the light passes through the movable section of the probe arms and passes underneath the probe arm bench to the fixed portion, mounted to the underside of the probe arm bench.
The fixed portion of each probe arm has several fold mirrors, a cold stop, a camera lens (an achromatic doublet of CaF$_{2}$ and S-FTM16 glass) and a flat fold `steering' mirror, which `steers' the light to each probe arm's beam combiner mirror.
The fixed portion of the probe arms (flat fold mirrors, cold stop, camera lenses and steering mirrors) make up what we call the aft system.

\begin{figure}[htp]
\centering
  \begin{minipage}[c]{0.48\textwidth}
  	\centering
	\includegraphics[height=0.48\textheight]{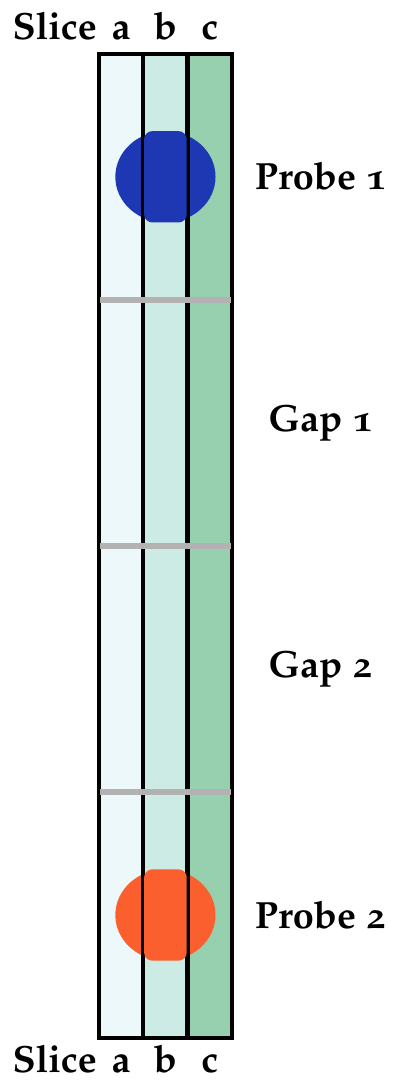}
	\caption{A schematic of a macro-slicer module.
   	  There are three mirrors which are oriented vertically in this representation.
   	  The two colored circles represent a 1\arcsec seeing disc from two probe arms.
	  Note that there are four more or less equal `slots' available on each of the slicer mirrors.
	  The green shading is constant for each mirror; the gray horizontal lines are there to guide the eye and do not represent physically separate mirrors.
	  Each macro-slicer is illuminated with two probes.
	  The two gaps are of equal size as the probe slots.
	  By leaving the middle $\sfrac{1}{2}$ of the macro-slicer module un-illuminated, we are able to interleave the images of the probes at the field mirrors as demonstrated in Figure~\ref{fig:macroSlicerInterleaving}.}
	\label{fig:macroSlicerSchematic}
  \end{minipage}
  \quad
  \begin{minipage}[c]{0.48\textwidth}
    \centering
    \includegraphics[height=0.48\textheight]{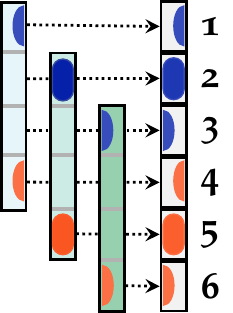}
    \caption{A schematic of how the images of the slicer mirrors are interleaved at the field mirror image plane.
    Note that the slicer mirrors are drawn separated and with an adjusted aspect ratio for compactness.
    The staggered position of the slicer mirrors reflects how the slices are geometrically rearranged with the aid of the pupil mirrors.
    On the far right is a set of field mirrors with numbers to the right.
    Each field mirror sees one illuminated 'slitlet' from the slicer module.
    The arrows show that each field mirror is illuminated by only one probe and zero, one, or two gaps.
    The field mirrors are 0\farcsec4 in width and \til3\farcsec7 in length.
    Not shown are the pupil mirrors, nor the de-magnification of the macro-slicer.}
   \label{fig:macroSlicerInterleaving}
  \end{minipage}
\end{figure}

The 12 beam combiner mirrors are simple flat fold mirrors in a monolithic array that mounts directly on the macro-slicer IFU bench and feed the 6 macro-slicer slicer modules (each one receiving two probes); there are a total of 18 slices for the 12 probes (see \S\ref{sec:numerology} for a discussion on the `numerology' of the macro-slicer).
The beam combiner mirrors are situated directly in front of the slicer modules and are co-planar with the slicer modules' longitudinal axis.
A selectable decker sits between the beam combiner mirror array and the slicer modules, and allows for light from one up to 12 simultaneous targets to be dispersed.
Each beam combiner feeds \sfrac{1}{4} of a slicer module -- note that only half of a slicer module is illuminated (see Figure~\ref{fig:macroSlicerSchematic}).
The aft system and beam combiner mirrors work together to bring the light from each probe arm to a focus at the slicer mirrors and were designed simultaneously with this in mind.

Each slicer module has three slices, and each slice has its own pupil mirror.
Like other advanced image slicers, the slicer and pupil mirrors act as a collimator-camera relay; we use a specific geometry to rearrange the field of view to squeeze our two dimensional field of view to only having one.
Each slice is illuminated with two probes at each end; in effect we create two `slitlets', each slice having two slitlets that are each \sfrac{1}{3} of a probe arm field of view.
This doubling-up of probes on each slicer module is demonstrated in Figures~\ref{fig:macroSlicerSchematic} and \ref{fig:macroSlicerInterleaving}.
The 6 slicer modules are mounted onto a common mount which is itself mounted to the IFU bench.
We discuss the optimization in \S\ref{sec:optimize}.

The slicer mirrors re-image the telescope pupil image onto the aptly-named pupil mirrors. 
The pupil mirrors create an image plane at the field mirrors; there are 6 field mirrors per slicer module.
All of the pupil mirrors have individual tilts to properly `steer' the re-imaged slices to the corresponding slitlets.

The pupil mirrors are split into 6 monolithic modules of three mirrors each; each slicer module (itself monolithically constructed) has a corresponding pupil module (as in Figure~\ref{fig:macroSlicerPics}).
Each pupil mirror module is mounted on a common mount, which is mounted to the IFU bench.

At its output, the macro-slicer feeds a total 36 field mirrors which are at an image plane.
These field mirrors generate the pseudo-slit for the spectrograph module, are unpowered, and have individual tilts to correct for telecentric errors introduced by the IFU optics.
Each slicer module feeds 6 field mirrors.

The macro-slicer IFU bench bolts to the MIRADAS bulkhead, which separates the spectrograph module from the pre-optics which include movable and fixed portions of the probe arms, the aft system, and the macro-slicer.
Each array, or block, of mirrors is designed with precision location pinholes and bolt holes for mounting.
This removes the onus of aligning some 84 mirrors in 14 separate arrays onto a single bench (note that this does not include the probe aft system with its additional 12 lens doublets and 48 mirrors!).
Each mirror array is machined out of RSA-6061 aluminum, with the optics being plated with electroless nickel, diamond-turned, and then gold-coated (as in FISICA\cite{fisica04a,fisica04b,fisica07} and FRIDA\cite{frida08,frida12}).
The optical mounts are also designed with precision location pinholes and bolt holes for mounting to the IFU bench.
The beam combiner and field mirror arrays pull double-duty as both mirror substrate and mounts to the IFU bench; the other arrays' mounts are machined out of 6061 aluminum.

\subsection{The Numerology of the Macro-Slicer}
\label{sec:numerology}
The `numerology' of the MIRADAS macro-slicer system was driven to match the optimal input to the spectrograph system (see Figure~\ref{fig:MI-SlitGeometry}).
As such, our design approach has been appropriately backwards, starting with the desired fixed output of the macro-slicer and keeping the more fluid input of the macro-slicer in mind.

The tilt of the spectra introduced by the geometry of the nearly Litthrow-mode echelle and cross-disperser gratings is mostly compensated by tilting the input (eg, the canonical stagger).
Without the canonical stagger, the `packing efficiency' of the spectra at the detector is only 50\% that of the staggered efficiency, a poor use of expensive detector real-estate!
The stagger comes in groups of 6 because there are two probes per slicer module.
We easily could have made 4 or even 6 probes per slicer, but the slicer modules 1) become correspondingly longer (4 probes is \til21 mm long), 2) the larger angles due to the size of the slicer increase the geometric aberrations introduced at the field mirrors by the pupil mirrors, and 3) changes the length of the canonical stagger, reducing the efficiency in which we fill the detector.
A more ideal solution would be a tilted pseudo-slit with no discrete stagger, but slicer mirrors work only when at an image plane, and if the field mirrors (which act as the input locations to the spectrograph) are at an image plane, the slicer mirrors must also be at an image plane; this is impossible to do if pupil mirrors are shared among slitlets while preserving AB nodding ability.

It also allows for relatively short slicer mirrors; a slicer mirror array made for all 12 probe arms would be \til200 mm long, as opposed to \til10 mm for the macro-slicer slicer mirrors.
This design choice allows for very low defocus due to mirror curvature in comparison to longer slicer mirrors as well as much easier fabrication.
However, the MIRADAS macro-slicer combines two probes as inputs for a single slicer module (see Figures~\ref{fig:macroSlicerSchematic} and \ref{fig:macroSlicerInterleaving} for a schematic of both the slicer module and the interleaving).

Another way to think about the macro-slicer approach is that we have made 12 advanced image slicers with three slices (the aforementioned slitlets) each; the slitlets are paired up and share a common center of curvature as well as a pupil mirror, but each has a unique field mirror.
Opto-mechanically, the paired slitlets are a single slice, which has one pupil mirror, and two field mirrors.
During the early design phase, we used the average position of the field mirrors in our calculations to ensure that the slitlet pairs really are indistinguishable from one slicer mirror.

Mechanically, the diamond-tipped tool that machines the slices is not double-edged, which requires making the slicer mirrors in two parts that do not meet in the middle of the mirror (due to the geometry of the cutting tool).
This means that the slicer mirrors will not be machined to a mirror finish in the middle of the slices, and we will have two `slicelet' mirrors sharing a common radius \& center of curvature and substrate.
For our purposes, however, the macro-slicer will not (and must not!) be illuminated in the middle, so the fact that the middle of our slicer mirrors aren't actually mirrors is moot.

\subsection{Comments on Optical Design Optimization}
\label{sec:optimize}
We used ZEMAX 13.2 for our optical modeling of the macro-slicer.
We started with a spreadsheet model which listed `nominal' slice, pupil, and field mirror positions with mirror tilts and radii of curvature calculated based on mirror size and position.
The `nominal' positions were chosen to satisfy the required output geometry of the macro-slicer.
As in Figure~\ref{fig:macroSlicerInterleaving}, each set of 6 field mirrors (corresponding to one macro-slicer module) pair naturally according to slice; field mirrors 1 \& 4, 2 \& 5, and 3 \& 6 share slices, so we used the average position of the field mirror pairs in our initial calculations. 
The outputs (average field mirror position) were kept fixed, as were the inputs (slices).
From there we modeled each slice--pupil--average-field mirror group separately, using the average position of the field mirror pair for each slice.

We built a merit function to further optimize each slice--pupil--average-field mirror group's performance, focusing on minimizing the pupil mirror size, the output spot size, and the output beam speed and direction; the latter was vital because the design of the spectrograph module of MIRADAS depends on a fixed output position and beam speed from the macro-slicer.
The pupil mirrors are individually optimized to minimize geometric aberrations via slight ‘pistoning’ along the vector from slicer mirror center to the nominal pupil mirror center (in other words we vary the distance between slice and pupil without changing the tilt angles of the slice) and allow the pupil mirror to have a slightly varying radius of curvature. 
This prioritizes optical performance as measured by the 50\% Encircled Energy Diameter (50\%EED, equivalent to the Full Width at Half-Maximum (FWHM) figure of merit commonly used in astronomy) at the plane orthogonal to the beam and co-located with the center of the field mirror.
We heavily weighted the output beam speed in our optimizer and kept the field mirror position fixed relative to the macro-slicer mirror, which means that our canonical stagger (Figure~\ref{fig:MI-SlitGeometry}) was `cooked in' to our model.

Once the design for each slice--pupil--average-field mirror group was complete and verified to not interfere mechanically with all other groups (particularly the pupil mirrors), we then modeled each slitlet--pupil--field mirror group, using the optimized slice--pupil--average-field model as a starting point.
We replaced the slice and average-field mirrors with their slitlet and field mirror replacements, and keeping slitlet and pupil mirror tilts fixed, as well as all mirror positions fixed.
The slitlet--pupil--field mirror model also includes the optical model of the spectrograph module.
The spectrograph module was designed well before the design of the macro-slicer was finalized, so we used the extant spot diagrams and spot centroids from the spectrograph module design as comparisons for our optimization scheme.
The merit function we used to optimize the slitlet's performance gave us the proper tilts to use for each of the field mirrors by heavily weighting the direction cosines of the outgoing beam from the field mirrors.
We also tested the performance using powered field mirrors, but found no meaningful improvement at the detector focal plane; given the added cost and difficulty in machining (and then measuring) powered field mirrors, we dismissed them.

\begin{figure}[ht]
\centering
  	\centering
	\includegraphics[width=0.5\textheight]{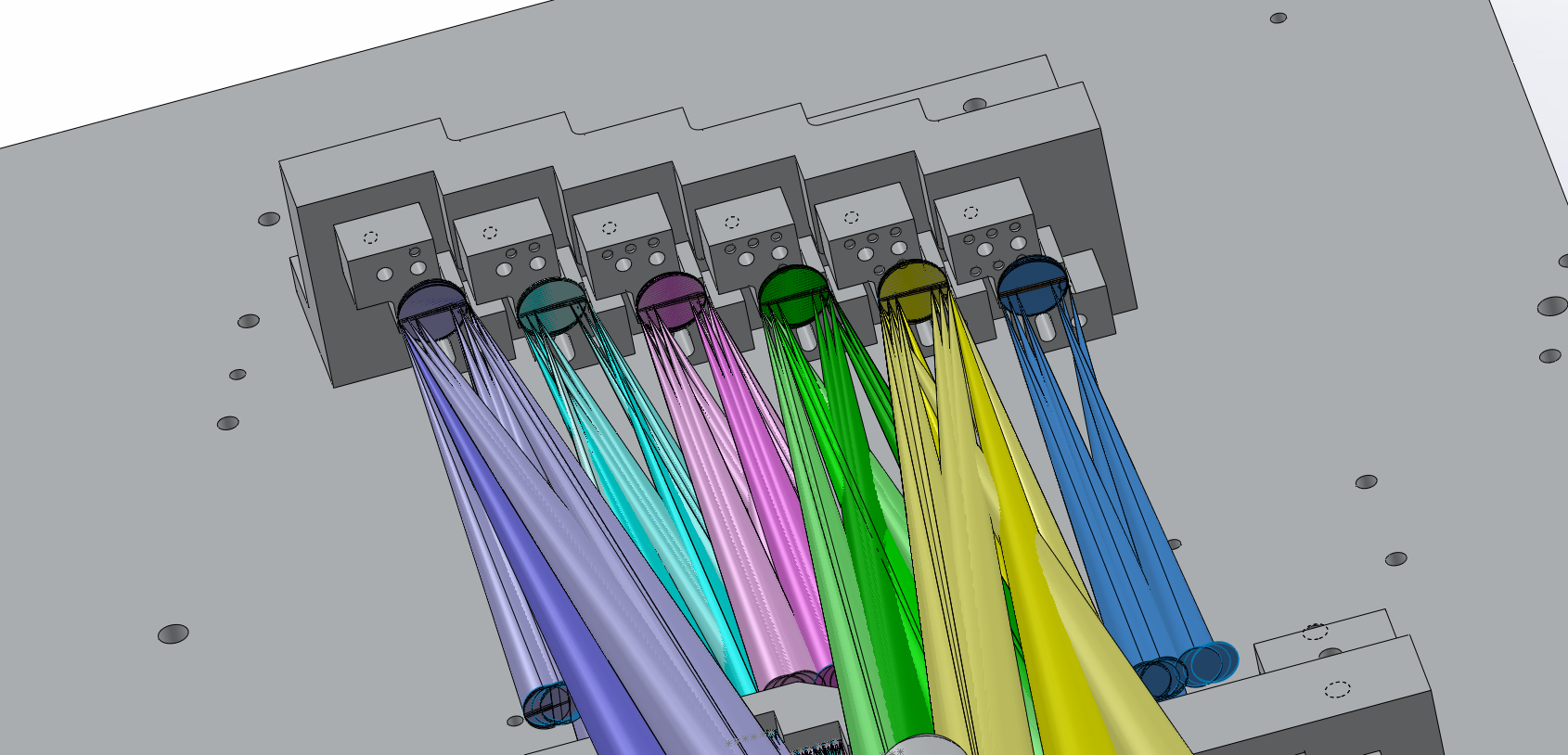}
	\caption{A view of the SolidWorks 3D CAD model of the macro-slicer, showing the 6 slicer units and the back end of the field mirrors; this model is for a 12 probe design.
	The imported beams for each of the 36 slitlets are also shown, each with similar colors for the 6 slicer modules.}
	\label{fig:beams}
  \end{figure}

\begin{figure}[ht]
\centering
  \begin{minipage}[l]{0.47\textwidth}
  	\centering
	\includegraphics[width=0.4\textheight]{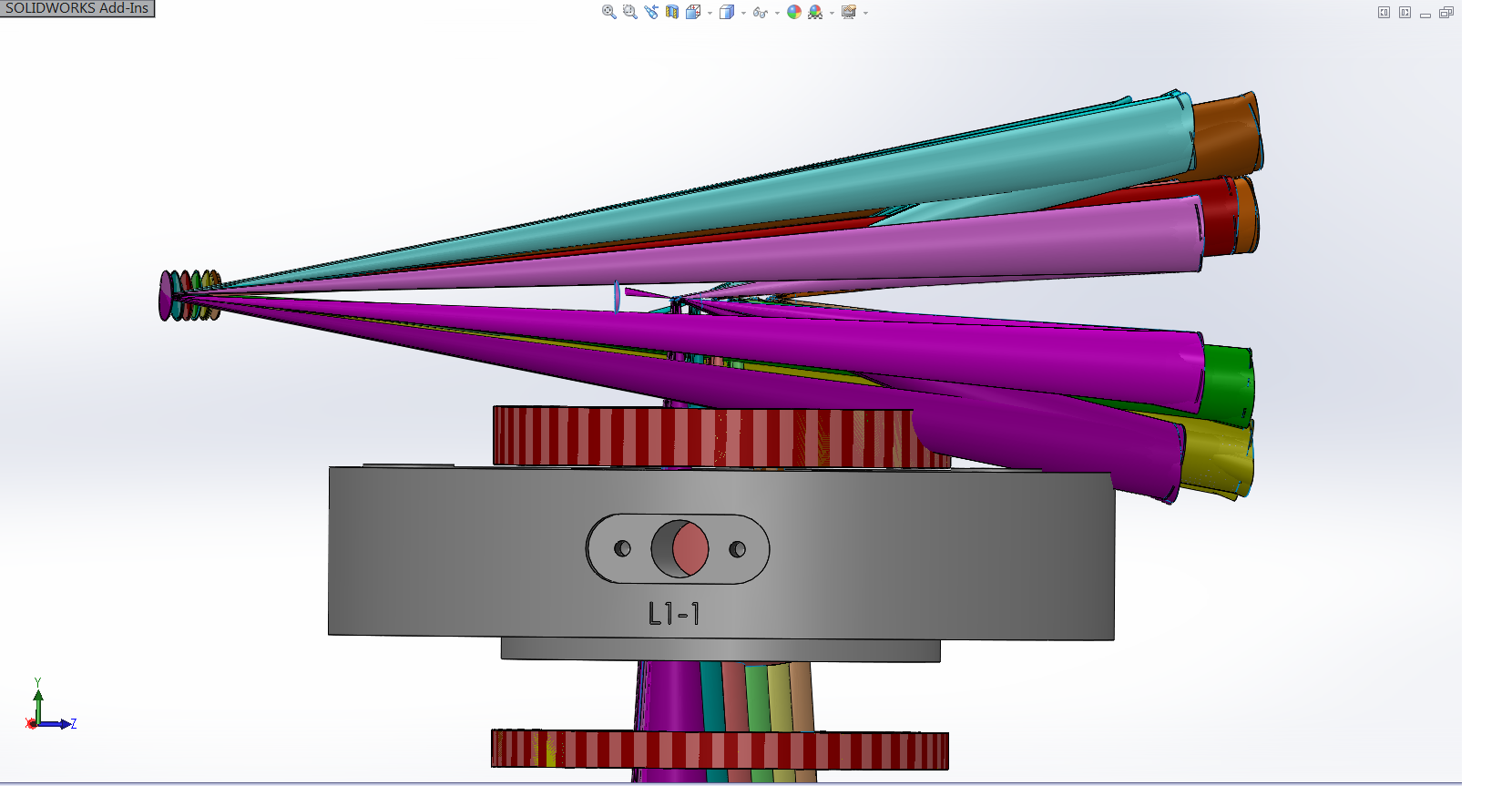}
  \end{minipage}
 \qquad 
  \begin{minipage}[r]{0.47\textwidth}
  	\centering
	\includegraphics[width=0.4\textheight]{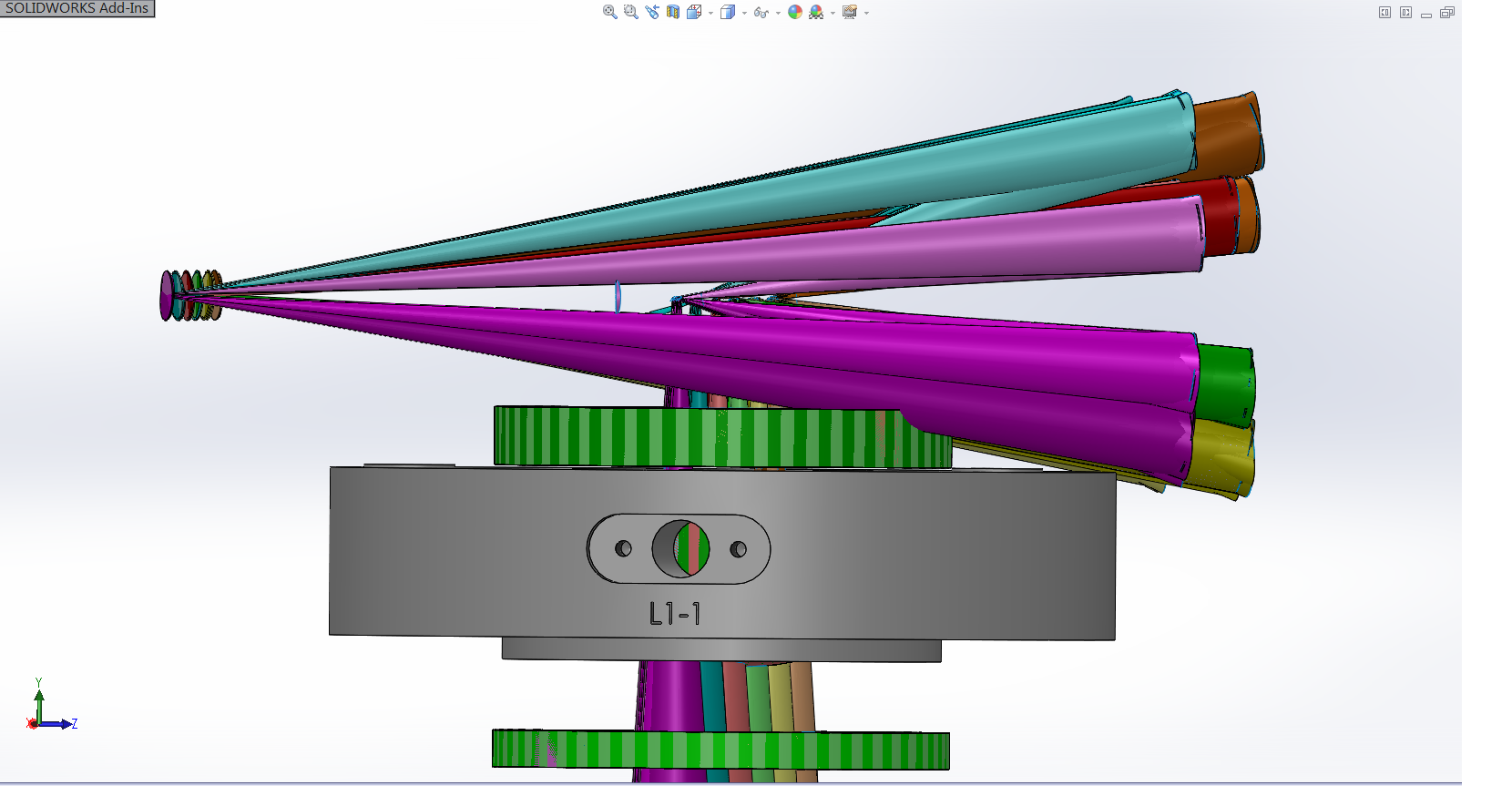}
  \end{minipage}
  \caption{A view of the beams exported from ZEMAX into SolidWorks.
    The colors for each slicer module are not the same as in Figure~\ref{fig:beams}.
	The collimator lens cell, L1, is shown; it is the first of 3 collimating lenses in the spectrograph module.\\
	\textbf{Left:} The lowest magenta pupil mirror is vignetted by the L1 cell.\\
	\textbf{Right:} With a slight adjustment of the angles of the slicer unit, we remove the vigetting entirely.}
  \label{fig:vignette}
\end{figure}

We also exported the beams from ZEMAX and compared them to our SolidWorks design, as in Figure~\ref{fig:beams}.
In doing so, we were able to check for vignetting and adjust the positions of our pupil mirrors as needed, as demonstrated in Figure~\ref{fig:vignette}.

This sophisticated modeling of the macro-slicer, \emph{slitlet by slitlet}, allowed us to verify that the macro-slicer performance is excellent and will not degrade the overall performance of MIRADAS.
We are able to make comparisons for any arbitrarily chosen order, wavelength, and input position between the spectrograph module design and the corresponding macro-slicer slitlet design.
Further, we were able to `dial in' the performance, with additional tolerance modeling showing that the spectrograph module can and does meet performance requirements as long as the output beam speed of the macro-slicer is within 1\% of the nominal f/7.10, easily achieved with our optimization scheme.

\subsection{Opto-mechanical Layout}
The mechanical design of the macro-slicer was tightly constrained by the working volume which is defined by the bulkhead, fore-ring, and probe arm bench (the fore-volume).
The fore-volume is further constrained by the requirement that the beam combiner mirrors, along with the aft system, keep a constant path length from the cold stop of the probe arms to the slicer mirror.
We wrote an iterating Python script to calculate positions and tilts for the steering mirrors and the tilts of the beam combiner mirrors.
It took into account the placement and exit angles of the beams as they left the probe arms, the sizes of the optics of the aft system, and the location of the beam combiner mirror array.
This allowed us to design both the macro-slicer beam combiner mirror array and the aft system in a coordinated fashion, and loosened the constraints of both by giving us more degrees of freedom in our design.

\section{Design Principles \& Trade Studies}
\subsection{Instrument Design and Heritage}
The MIRADAS instrumentation team takes advantage of a wealth of previous experience with cryogenic instrumentation.
We use the `bolt-and-go' approach for alignment and monolithic construction combined with diamond-turning for our mirror arrays.

\begin{figure}[htp]
  \begin{minipage}[c]{0.45\textwidth}  
    \centering
    \includegraphics[width=0.95\textwidth]{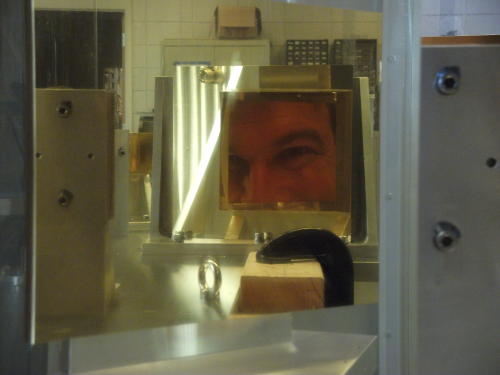}
    \caption{The view from the entrance side of CIRCE with one of the co-authors in focus at the detector position.
      Note that the mirrors are gold-coated and diamond-turned aluminum.
      The mounts for several of the mirrors can be seen, along with the precision location pinholes.
      Because there are no refractive optics, the team was able to accurately test CIRCE at room temperature and pressure with visible-light techniques.}
    \label{fig:CIRCE-Steve}
  \end{minipage}
  \qquad
  \begin{minipage}[c]{0.45\textwidth}  
    \centering
    \includegraphics[width=0.95\textwidth]{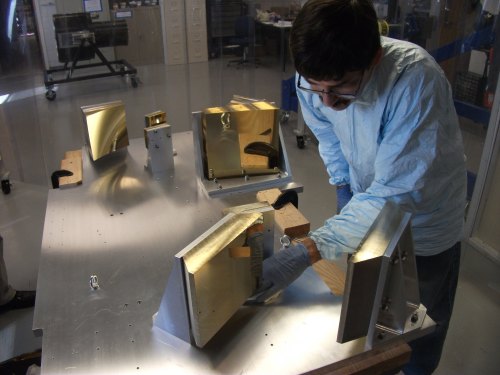}
    \caption{Overhead view of the CIRCE bench.
      Each optic is gold-coated and diamond-turned aluminum.
      The mirrors and mounts were machined from the same billet of aluminum, while the bench was machined out of the same alloy; this guarantees precise CTE matching.
      Also featured is Dr. Nick Raines inspecting one of the flat fold mirrors.}%
    \label{fig:CIRCE-Nick}
  \end{minipage}  
\end{figure}

\begin{figure}[htp]
  \centering
  \begin{minipage}[c]{0.45\textwidth}  
    \centering
    \includegraphics[width=0.95\textwidth]{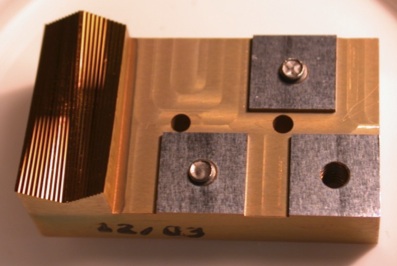}
    \caption{The FISICA slicer mirror array. 
      Note the pin holes, tapped bolt holes, and mounting pads are all machined into the mirror substrate.
      Each slice has independent tilt angles and are equally powered.
      The mirrors are 9.1 mm long and 0.195 mm wide.}
    \label{fig:FISICA-SM}
  \end{minipage}
  \qquad
  \begin{minipage}[c]{0.45\textwidth}  
    \centering
    \includegraphics[width=0.95\textwidth]{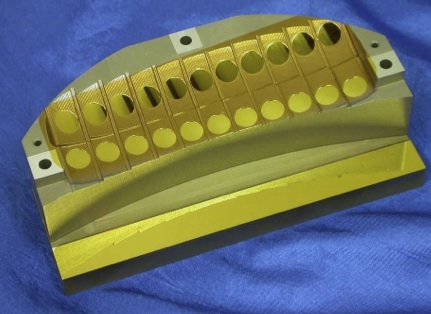}
    \caption{The FISICA pupil mirror array (as well as an integrated flat mirror at lower right). 
      Note the pin holes, tapped bolt holes, and mounting pads are all machined into the mirror substrate.
      Each pupil mirror is independently tipped, tilted, and powered.
      The mirrors are approximately 10 mm in diameter.}
    \label{fig:FISICA-PM}
  \end{minipage}
\end{figure}

The macro-slicer design builds on previous successful designs by our group such as FISICA, CIRCE\cite{circe04,circe14}, and FRIDA.
FISICA is a seeing-limited Content-style advanced image slicer and was designed to work with the FLAMINGOS-1\cite{flamingos98,flamingos03} instrument.
CIRCE has large, diamond-turned mirrors (Figures~\ref{fig:CIRCE-Steve} and \ref{fig:CIRCE-Nick} show CIRCE's optical bench with mirrors installed), two of which are 14th order even aspherical surfaces.
FRIDA is a diffraction-limited Content-style advanced image slicer designed for GTC.
All three are manufactured in monolithic blocks, where each set of mirrors is machined out of the same billet of aluminum to ensure exquisite coefficient of thermal expansion (CTE) matching.
FISICA's slicer mirror arrays (the slicer, pupil, and field mirror arrays) are machined into three blocks of metal, which puts the onus of alignment in the design (and machining) each mirror array; Figures~\ref{fig:FISICA-SM} and \ref{fig:FISICA-PM} show the FISICA slicer mirror and pupil mirror arrays.
All three instruments use our group's innovative `bolt-and-go' approach, with each mirror and mount machined with kinematic mounting pads, bolt- and pin-holes; this allows all of the grueling alignment work to be done in the design phase, \emph{not} in the assembly phase.
As a result, the precise alignment of these complex optical systems a side-effect of assembly.

\subsection{Trade Studies}
\label{sec:tradeStudies}
The macro-slicer concept (breaking up the slicer mirror in to several units and interleaving the probes at the field mirrors) is a new twist on the advanced image slicer design.
We looked at a large swathe of design space, a portion of which is described below.
The preliminary design called for three pupil mirrors and an array of ten macro-slicers, each with two sets of slicer mirrors (the MIRADAS preliminary design also called for up to 20 probes).
During the final design, the total number of probe arms was capped at 12 (reducing the mass, and cost of the gratings and spectrograph camera lenses).
This allowed us to simplify the design of the slicer mirrors while still utilizing the macro-slicer concept.

We examine some of the trade studies we made in the following sub-sections.

\subsubsection{Number of slicer modules}
Early designs for MIRADAS called for up to 20 probe arms with a correspondingly larger psuedo-slit (\til300 mm at the macro-slicer input and \til150 mm at the macro-slicer output).
In order to keep the performance as high as possible (eg, the \fiftyeed~as small as possible), the macro-slicer modules were designed to be in two rows and to share a common set of three pupil mirrors to six pupil mirrors.
However, with the descoping of number of probe arms from 20 to 12 (which also reduced the diameter, and thus cost, complexity and risk, of the gratings and spectrograph camera lenses), we found we could greatly improve the performance of all probe arms if we used one pupil mirror per slice.
This allowed us to `tune' the pupil mirror to maximize the performance on a slice-by-slice basis.

\subsubsection{Number of pupil mirrors}
The initial design called for three pupil mirrors for all slicer modules (one mirror per 'group' of slices).
However, we found that the performance of most slices would be worsened as a result, so we increased the number of pupil mirrors to one per slice.

Later, we examined using two slices from adjacent slicer modules per pupil mirror in order to reduce the number of pupil mirrors.
While the spacing of the slicer modules could be adjusted to give the proper spacing of slitlets at the location of the field mirrors, the canonical stagger is not preserved.
Even if we were to create a simple pseudo-slit with no stagger, we were unable to get proper positioning of the images of the slitlets at the field mirror positions, nor were we able to tune the performance of the slitlets without significantly compromising half of the slitlets' image quality.

\subsubsection{Powering the field mirrors}
We looked at powering the field mirrors to improve the telecentric performance of the macro-slicer.
However, we found that the telecentric was within requirements even for the worst case  without having to power the field mirrors.
Additionally, having to machine such small powered mirrors would have increased cost.

\subsubsection{Using off-axis parabolic mirrors instead of spherical mirrors}
The motivation for using Off-Axis Parabolic mirrors (OAPs) is that OAPs are very good in comparison to spherical mirrors at taking off-axis collimated beams and delivering them to a focus, which suffer from spherical aberration that increases strongly with the degree of off-axis.
What we found was that the OAPs only slightly improve performance for our pupil mirrors.
When we allowed the conic constant to vary (spherical mirrors have a conic constant of 0; parabolas -1; values $<$-1 are hyperbolic), we found that the best improvement in spot size (\fiftyeed) was with a conic constant of \til~-8 -- in other words, strongly hyperbolic.
The performance gain was not high enough to warrant the increased cost in machining.
Additionally, to our knowledge, OAP mirrors, let alone off-axis hyperbolic mirrors, have not been demonstrated to work as pupil mirrors in an advanced image slicer.

\subsubsection{Varying the radii of curvature of the slicer and pupil mirrors}
We also looked at allowing the slicer and pupil mirror radii of curvature to vary.
There was not much gain to be had by varying the slicer mirrors (also, the increased cost due to re-tooling), so we concentrated on varying the pupil mirrors.
We found that varying the pupil mirrors radii of curvature increased performance, especially after we allowed the spacing between the slicer and pupil mirrors to vary as well.
In effect, we were able to `dial in' the nominal values for spot centroid position in our optimization and focus on minimizing spot size and wavefront RMS error.
We selected this method of optimization for our final design, as discussed in \S\ref{sec:optimize}.

\section{Expected Performance}
\label{sec:performance}
\begin{figure}[htp]
  \centering
  \begin{minipage}[c]{0.23\textwidth}  
    \centering
    \includegraphics[width=\textwidth]{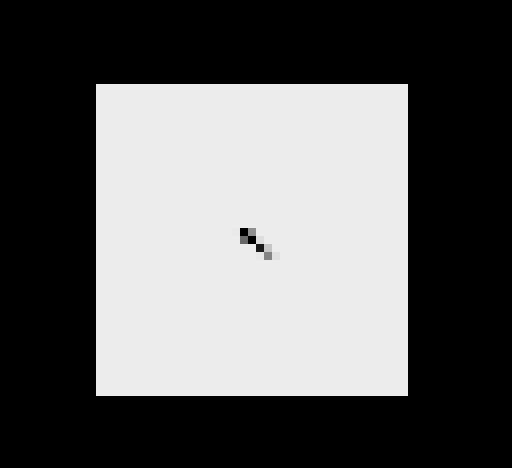}
  \end{minipage}
  ~
  \begin{minipage}[c]{0.23\textwidth}  
    \centering
    \includegraphics[width=\textwidth]{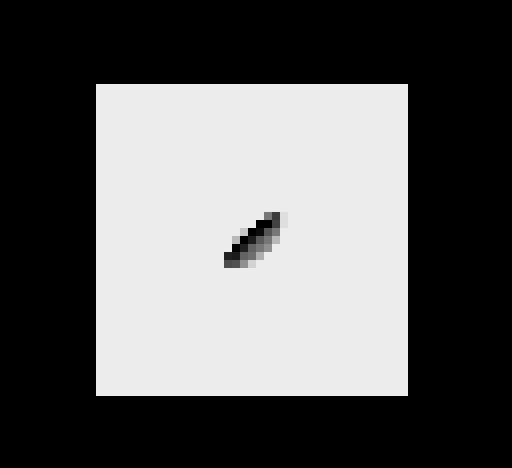}
  \end{minipage}
  ~
  \begin{minipage}[c]{0.23\textwidth}  
    \centering
    \includegraphics[width=\textwidth]{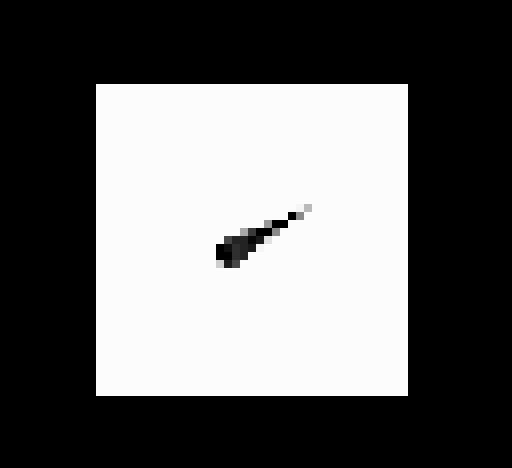}
  \end{minipage}
  ~
  \begin{minipage}[c]{0.23\textwidth}  
    \centering
    \includegraphics[width=\textwidth]{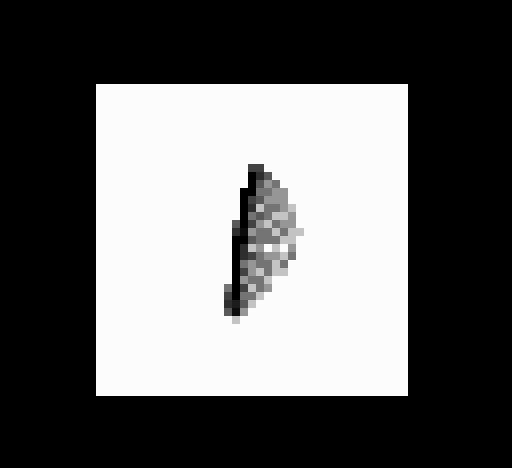}
  \end{minipage}
  \captionof{figure}{\small{
  Four typical synthetic PSFs are shown projected onto the detector focal plane.
  The pixel scale is 4.5$\mu$m, or 0.25 science detector pixels.
  The PSFs, left to right, are from probes 6, 7, 9, and 2.}
  }
  \label{fig:syntheticPSFs}
\end{figure}

We are able to simulate the end-to-end performance of the MIRADAS instrument by use of the MIRADAS Data Simulator (MIRA-DaS).
MIRA-DaS comes in two parts; the first (static) stage uses full-system ZEMAX models (one per slitlet, or 6 per probe) to calculate the instrumental response as a function of wavelength and position at the probe tip.
We generate hundreds of thousands of PSFs at the detector plane; there are 5 positions, one for the center and each corner, of any given slitlet.
The PSFs are wavelength-dependent as well as field position (which slitlet and where on each slitlet).
Each PSF is over-sampled on a pixel grid that is finer than the MIRADAS detector pixel size; Figure~\ref{fig:syntheticPSFs} shows an example of four synthetic PSFs.
Overall, each mode (multi-object, single-object short wavelength bandpass, and single-object long wavelength bandpass) has over 20~000 generated PSFs, each with thousands of individual rays traced.

This huge PSF library makes up a well-sampled grid, which MIRA-DaS then uses to generate synthetic spectra, given an input template spectrum.
MIRA-DaS steps across in wavelength space and source location, using the intensity from the input spectrum to generate emission or absorption features.
Each slitlet of the macro-slicer is treated individually to allow for ease of computation and allows us to perform some degree of parallelization.
The steps in wavelength space are arbitrarily small; typically we use a wavelength step of 10\AA, which allows us to super-sample each detector spaxel before final binning occurs.
The PSF generated by the probe arm can be described as a typical Gaussian (only a slight lie, but a not unreasonable one), but the PSF is decidedly non-Gaussian after it passes through the macro-slicer optics.

The performance of the macro-slicer (and, MIRADAS as a whole) is, quite simply, exquisite.
We find that the resolution in each of the spectral orders to exceed by \til10\% the required $R\approx20,000$ out to the edges of each order.
The throughput of MIRADAS is a few percent at any given wavelength, on par with any other high-resolution spectrograph, most of which lack the multi-object ability of MIRADAS!

\section{Closing Remarks}
\label{sec:closing}
We have described the MIRADAS macro-slicer, an innovative take on the advanced image slicer concept.
The macro-slicer design fulfills several required tasks: 
\begin{enumerate}
	\item Our slitlet-by-slitlet design approach allows for direct comparisons between the spectrograph module design and the macro-slicer+spectrograph module design.
	\item All twelve probe arms are `funneled' into a common pseudo-slit, which is staggered to counteract the innate tilt introduced by the cross-dispersed echelle geometry, maximizing the use of the detector area.
	\item Delivers seeing-invariant performance, mitigating the need for a selectable decker.
	\item The macro-slicer modules allow for a small amount of spatial information to be recovered from extended objects in good seeing.
	\item On-slit AB nodding capability:
	By ensuring the macro-slicer assembly is telecentric, we guarantee the ability of MIRADAS to use AB nodding for removal of telluric lines during data processing.
\end{enumerate}

\acknowledgements
The authors acknowledge the support of the University of Florida, the MIRADAS Consortium, the Gran Telescopio Canarias, and the Fundacion por la Preservacion del PSF.
We gratefully acknowledge Salvador Cuevas (UNAM) for his helpful advice and insights in the design and implementation of advanced image slicers, in particular how to do so with ZEMAX.

\bibliography{MacroSlicer}
\bibliographystyle{spiebib}
\end{document}

%% file: opticalPrescription.tex


\small{
\begin{tabularx}{0.9\textwidth}{cccccccc}
\toprule
	&  &	Radius of	& f-speed	&	& Diameter/	&	& 	\\
Component	& \# Mirrors	&	Curvature	& (outgoing)	& Plate Scale	&  Length	& Width	& Thickness \\
	&	&  [mm] & [unitless]	& [\sfrac{\arcsec}{mm}]	& [mm]	& [mm]	& [mm]	\\
\midrule
Beam combiner	& 12	& $\infty$	& 14.15	& ---	& 10	& 10	& 50	\\
Slicer	& 18	& 450	& 14.15	& 1.4575	& 10.365	& 0.2675	& 225	\\
Pupil	& 18	& 225	& 7.10	& ---	& 16	& ---	& 115	\\
Field	& 36	& $\infty$	& 7.10	& 2.9050	& 1.29	& 0.5	& --- \\
\bottomrule
\end{tabularx}
} 